\documentclass[conference]{IEEEtran}
\IEEEoverridecommandlockouts
\usepackage{cite}
\usepackage{amsmath,amssymb,amsfonts}
\usepackage{algorithm}
\usepackage{algcompatible}
\usepackage{float} 
\usepackage{graphicx}
\usepackage{textcomp}
\usepackage{xcolor}

\makeatletter
\newenvironment{breakablealgorithm}
  {
   \begin{center}
     \refstepcounter{algorithm}
     \hrule height.8pt depth0pt \kern2pt
     \renewcommand{\caption}[2][\relax]{
       {\raggedright\textbf{\ALG@name~\thealgorithm} ##2\par}%
       \ifx\relax##1\relax 
         \addcontentsline{loa}{algorithm}{\protect\numberline{\thealgorithm}##2}%
       \else 
         \addcontentsline{loa}{algorithm}{\protect\numberline{\thealgorithm}##1}%
       \fi
       \kern2pt\hrule\kern2pt
     }
  }{
     \kern2pt\hrule\relax
   \end{center}
  }
\makeatother

\def\BibTeX{{\rm B\kern-.05em{\sc i\kern-.025em b}\kern-.08em
    T\kern-.1667em\lower.7ex\hbox{E}\kern-.125emX}}
\begin{document}

\title{Rough Set improved Therapy-Based Metaverse Assisting System}

\author{
\IEEEauthorblockN{1\textsuperscript{st*} Jin Cao}
\IEEEauthorblockA{
   \textit{Independent Researcher} \\
   \textit{Johns Hopkins University}\\
   Baltimore, MD, 21218, USA \\
   caojinscholar@gmail.com}
\and
\IEEEauthorblockN{2\textsuperscript{nd} Yanhui Jiang}
\IEEEauthorblockA{
   \textit{Department of Computer Science} \\
   \textit{University College London (UCL)}\\
   London, WC1E 6BT, UK\\
   yanhui.jiang.23@ucl.ac.uk}
\and
\IEEEauthorblockN{2\textsuperscript{nd} Chang Yu}
\IEEEauthorblockA{
   \textit{Independent Researcher} \\
   \textit{Northeastern University}\\
   Boston, MA, 02115, USA \\
   chang.yu@northeastern.edu}
\and
\IEEEauthorblockN{\hspace{25mm}3\textsuperscript{rd} Feiwei Qin}
\IEEEauthorblockA{
   \hspace{25mm}\textit{School of Computer Science and Technology} \\
   \hspace{25mm}\textit{Hangzhou Dianzi University}\\
   \hspace{25mm}Hangzhou, Zhejiang, 310005, China \\
   \hspace{25mm}qinfeiwei@hdu.edu.cn}
\and
\IEEEauthorblockN{4\textsuperscript{th} Zekun Jiang}
\IEEEauthorblockA{
   \textit{College of Computer Science} \\
   \textit{Sichuan University}\\
   Chengdu, Sichuan, 610065, China \\
   zekun\_jiang@163.com}
}

\maketitle

\begin{abstract}
Chronic neck and shoulder pain (CNSP) is a major global public health issue. Traditional treatments like physiotherapy and rehabilitation have drawbacks, including high costs, low precision, and user discomfort. This paper presents an interactive system based on Cognitive Therapy Theory (CBT) for CNSP treatment. The system includes a pain detection module using EMG and IMU to monitor pain and optimize data with Rough Set theory, and a cognitive therapy module that processes this data further for CBT-based interventions, including massage and heating therapy. An experimental plan is outlined to evaluate the system's effectiveness and performance. The goal is to create an accessible device for CNSP therapy. Additionally, the paper explores the system's application in a metaverse environment, enhancing treatment immersion and personalization. The metaverse platform simulates treatment environments and responds to real-time patient data, allowing for continuous monitoring and adjustment of treatment plans.
\end{abstract}


\section{Introduction}
Chronic neck and shoulder pain (CNSP) is a pervasive challenge affecting public health and socioeconomic well-being globally \cite{b40}, adversely impacting individuals' quality of life and exerting broader effects on societal productivity and healthcare expenditure \cite{b40}. Current CNSP treatment relies on conventional approaches like physiotherapy and medical rehabilitation, aligning with prevailing practices for chronic pain management \cite{b33}.

While traditional CNSP treatments are widely used, significant gaps exist in their ability to offer efficient, interactive, and user-friendly solutions \cite{b39}. In addition, these treatments suffer from drawbacks such as high costs and neglect of the user's psychological experience \cite{b2}. Also, on the technical side, diagnostic image detection\cite{b51, b52} helps the additional neck or shoulder pain detection from the imaging side. Conversely, Cognitive Behavioural Therapy (CBT) offers an inclusive approach, considering psychological aspects and actively guiding CNSP management. In the three decades since CBT was first applied to chronic pain, much research has confirmed the importance of cognitive and behavioral processes in how individuals adapt to chronic pain \cite{b18}. Growing interest exists in integrating cognitive therapy techniques to address behavioral dysfunction \cite{b1}. 

This study introduces a novel CBT-based approach as a complementary therapeutic tool for CNSP—an interactive system supporting non-professional treatment and alleviating mental discomfort \cite{b18}. We will optimize the raw data collected by two Arduino modules through Rough Set theory\cite{b43}, and we have also designed subsequent validation experiments and plans to assess the effectiveness and performance of the interactive system, including quantitative interviews and qualitative surveys. 

The rest of this paper is organized as follows. Section II reviews the related work on CNSP treatments and advances in cognitive therapy. Section III presents the equations and algorithms used to process and analyze the EMG data, utilizing rough set theory for enhanced accuracy. Section IV introduces a novel CBT-based system for CNSP treatment, detailing both the pain detection and cognitive therapy modules. Section V outlines the evaluation plan, including experimental setup and methodologies for qualitative and quantitative analysis. Finally, Section VI discusses the expected outcomes of the interventions, the integration of treatment within the metaverse, and the implications for future research.

\section{Related Work}

\subsection{Therapeutic Interventions for CNSP}

As a chronic and recurrent condition, CNSP is usually characterized by stiffness and muscle pain, a common musculoskeletal problem associated with a common disease \cite{b25}. Most commonly, physiotherapy, medication, and acupuncture provide only temporary relief, often leading to recurrence \cite{b15}. Limitations such as lack of precision and recurrent painful episodes of conventional treatments become apparent \cite{b34}. In addition, researchers have explored the use of High-intensity laser (HIL) technology therapy \cite{b30}. However, these treatments ignore pain triggers in response to intervention \cite{b28}. This has led to some patients reporting physical and psychological discomfort during the procedure, compromising compliance with the treatment plan \cite{b28}.

It is critical to understand the patient experience in CNSP treatment [17]. Existing pain-free options often lack seamless integration with mental health, with delayed feedback and inadequate psychological healing or response \cite{b5}. Our proposed solution addresses these issues through timely detection and attention to the users’ mental health.

\subsection{Cognitive Behavioural Therapy}
Our review examines the development of CBT since its inception as a popular therapy in the 1970s, which evolved into multiple treatment options in the 2000s \cite{b7}. Over the past 60 years, the scientific understanding of pain and the development of CBT have led to its use as an adjunctive treatment for chronic pain, either alone or in combination with rehabilitation \cite{b1,b17}. However, challenges remain due to the predominantly theoretical nature of CBT, the need for more viable, practical products, and the fact that standardized protocols still need revision, leading to challenges in determining clinical relevance \cite{b23,b29}. We also learned that using new tech for the user brain's understanding \cite{b50}, or enhanced large language/text models directly input user's needs\cite{b53,b54}, might also help our research for better treatment by new technologies.

In summary, although modern physical and medical device treatments for CNSP have research support, the evidence taken together suggests that existing biological and pharmacological interventions may cause pain or psychological distress in users \cite{b8,b26}. Furthermore, CBT-based treatment of chronic pain, although practical, still faces challenges, such as an emphasis on theory and a lack of tangible products \cite{b21}. Therefore, our study will improve the weaknesses in the user interaction and product implementation dimensions.

\section{EQUATIONS AND ALGORITHM}

This equation first processes the samples with Arduino to compute the EMG data, and then screens meaningful data using rough set theory and importance. To Enhance EMG Detection of Objective Pain Values Using Rough Set Theory. The incorporation of rough set theory helps in handling uncertainties and enhancing the precision of the pain detection algorithm.

\subsection{Equations 1: EMG Processing with Arduino}

Input: Raw EMG sample data

Output: Processed EMG data

\subsubsection{Initialize Arduino System}

Set up the Arduino analog input pins and initialize serial communication.

\subsubsection{Data Collection}

For each sample, read the EMG signal value \( \text{samples}[i] \).

\subsubsection{Data Processing}

Initialize minimum value \( \text{minValue} \) and maximum value \( \text{maxValue} \).

(3a). Calculate the total sum of samples \( \text{sum} \), which calculates the total sum of all sample values (\(\text{sum}\)). The expression \(\sum_{i=1}^{\text{SAMPLE\_COUNT}} \text{samples}[i]\) represents the summation of all sample values, where \(\text{SAMPLE\_COUNT}\) is the total number of samples:
     \[
     \text{sum} = \sum_{i=1}^{\text{SAMPLE\_COUNT}} \text{samples}[i]
     \]

(3b). Calculate the average value, which calculates the average value of the adjusted dataset. The numerator is the adjusted sum (\(\text{sumAdjusted}\)), and the denominator is the total number of samples minus two (because we have removed the two extreme values: the minimum and maximum). This average value (\(\text{AverageValue}\)) is more representative of the central tendency of the dataset since it excludes values that could disproportionately affect the result:
     \[
     \text{AverageValue} = \frac{\text{sumAdjusted}}{\text{SAMPLE\_COUNT} - 2}
     \]

\subsubsection{Data Logging and Monitoring}

Output the calculated average value \( \text{AverageValue} \) to the serial port and log it.

\subsubsection{Filtering and Amplification}
 
Apply filtering to remove noise from the collected EMG signals.
Amplify the filtered EMG signals to enhance signal strength.

\subsubsection{Feature Extraction}
 
Calculate the Mean Value of the EMG Signals \( \text{meanEMG} \), here we calculate the mean (average) value of the amplified EMG signals. In the formula, \( n \) represents the total number of EMG signal samples, and \(\text{amplifiedEMG}[i]\) represents the amplified EMG value of the \( i \)-th sample. By summing all the sample values and dividing by the total number of samples \( n \), the mean value of the EMG signals is obtained:
     \[
     \text{meanEMG} = \frac{1}{n} \sum_{i=1}^{n} \text{amplifiedEMG}[i]
     \]

\subsubsection{Analysis and Feedback}
 
Analyze the processed EMG data and provide feedback on muscle activity.

\subsection{Equations 2: Screening Meaningful Data Using Rough Set Theory and Importance}

Input: Processed EMG data

Output: Screened meaningful data

\subsubsection{Data Processing and Normalization}
 
Initialize minimum value \( \text{minValue} \) and maximum value \( \text{maxValue} \) from the processed samples from Equations 1.
   
\subsubsection{Incorporate Rough Set Theory}
 
Define the information system \( S = (U, A, V, f) \), where:
  \( U \) is the set of normalized samples.
  \( A \) is the set of attributes (e.g., normalized EMG features).
  \( V \) is the set of attribute values.
  \( f: U \times A \rightarrow V \) is the attribute function assigning values to each sample.
Determine the indiscernibility relation \( IND(B) \), this indicates that for each pair of samples \( (x, y) \), if for all attributes \( a \in B \), the attribute values \( f(x, a) \) and \( f(y, a) \) are equal, then \( (x, y) \) belongs to the indiscernibility relation \( IND(B) \):
     \[
     IND(B) = \{(x, y) \in U \times U \mid \forall a \in B, f(x, a) = f(y, a)\}
     \]

\subsubsection{Calculate Dependence and Importance}
 
Calculate the dependence \( \rho_a \) of each attribute \( a \in A \) on the target set \( X \), this calculates the dependence \( \rho_a \) of each attribute \( a \in A \) on the target set \( X \), where \( POS_B(X) \) is the number of samples in the positive region determined by attributes \( B \), and \( U \) is the total number of samples:
     \[
     \rho_a = \frac{|POS_B(X)|}{|U|}
     \]
Calculate the importance \( \gamma_a \) of each attribute \( a \in A \), this calculates the importance \( \gamma_a \) of attribute \( a \), where \( POS_{B-\{a\}}(X) \) is the number of samples in the positive region determined by the remaining attributes \( B-\{a\} \), excluding \( a \):
     \[
     \gamma_a = \frac{|POS_B(X)| - |POS_{B-\{a\}}(X)|}{|U|}
     \]
Combine dependence and importance to compute the overall weight \( \omega_a \) for each attribute, where \( \alpha \) and \( \beta \) are weight parameters, and \( \alpha + \beta = 1 \):
     \[
     \omega_a = \alpha \cdot \rho_a + \beta \cdot \gamma_a
     \]
Normalize the overall weights, this normalizes the overall weight \( \omega_a \) of each attribute, ensuring that the sum of all weights equals 1:
     \[
     \omega_a' = \frac{\omega_a}{\sum_{a \in A} \omega_a}
     \]

\subsubsection{Output Screened Samples}
 
   Output the screened sample set \( X' \), consisting of EMG data that have been refined through rough set theory to identify the most meaningful samples. These screened EMG data provide a clearer and more accurate depiction of the user's neck and shoulder pain status, allowing for more precise diagnosis and targeted treatment strategies.

\subsection{Algorithm 1: EMG Processing with Arduino}

Algorithm 1: In our research, to objectively evaluate the functionality of the muscular and nervous systems, we developed and implemented an EMG-based pain detection module. The system outputs this average value to the serial monitor. This average value represents the muscle pain level over a specific time interval, providing a reliable basis for the Cognitive Behavioral Therapy (CBT) assistive device to deliver targeted treatment. This process not only enhances the objectivity and accuracy of pain detection but also facilitates real-time feedback for CBT-based therapeutic interventions.

Here the input data is Raw EMG sample data, and the Output data is Processed EMG data by Adruino.

\begin{breakablealgorithm}
\caption{EMG Processing with Arduino}
\begin{algorithmic}[1]
\STATE Initialize variables: \texttt{samples}, \texttt{minValue}, \texttt{maxValue}, \texttt{sum}, \texttt{sumAdjusted}, \texttt{AverageValue}
\STATE Collect EMG samples into \texttt{samples} array
\STATE Set \texttt{minValue} and \texttt{maxValue} to the first sample value
\FOR{each sample in \texttt{samples}}
    \STATE Update \texttt{minValue} and \texttt{maxValue}
    \STATE Add sample to \texttt{sum}
\ENDFOR
\STATE Calculate \texttt{sumAdjusted} by subtracting \texttt{minValue} and \texttt{maxValue} from \texttt{sum}
\STATE Calculate average value:
\[
\text{AverageValue} = \frac{\text{sumAdjusted}}{\text{SAMPLE\_COUNT} - 2}
\]
\STATE Output \texttt{AverageValue} to serial monitor
\STATE (Optional) Apply filtering to remove noise and amplify the filtered EMG signals
\STATE Calculate mean and peak values of the EMG signals:
\[
\text{meanEMG} = \frac{1}{n} \sum_{i=1}^{n} \text{amplifiedEMG}[i]
\]
\[
\text{peakEMG} = \max_{i=1}^{n} \text{amplifiedEMG}[i]
\]
\STATE Analyze processed EMG data and provide feedback on muscle activity.
\end{algorithmic}
\end{breakablealgorithm}

\subsection{Algorithm 2: Screening Meaningful Data Using Rough Set Theory and Importance}

Algorithm 2 starts by normalizing the processed EMG data. This involves adjusting the data values so they fit within a consistent scale, making it easier to compare different samples.

The final output is a set of meaningful data, which highlights significant characteristics in the EMG data. This refined data set is particularly useful for understanding the user's neck and shoulder pain status, as it emphasizes the most relevant features from the EMG readings.

\begin{breakablealgorithm}
\caption{Screening Meaningful Data Using Rough Set Theory and Importance}
\begin{algorithmic}[1]
\STATE Initialize variables: \texttt{minValue}, \texttt{maxValue}, \texttt{normalizedSamples}, \texttt{dependence}, \texttt{importance}, \texttt{overallWeight}
\STATE Calculate \texttt{minValue} and \texttt{maxValue} of processed EMG data
\FOR{each sample in \texttt{processedSamples}}
    \STATE Normalize the data:
    \[
    \text{normalizedSamples}[i] = \frac{\text{processedSamples}[i] - \text{minValue}}{\text{maxValue} - \text{minValue}}
    \]
\ENDFOR
\STATE Define the information system $S = (U, A, V, f)$ and determine the indiscernibility relation $IND(B)$:
\[
IND(B) = \{(x, y) \in U \times U \mid \forall a \in B, f(x, a) = f(y, a)\}
\]
\FOR{each attribute $a \in A$}
    \STATE Calculate dependence $\rho_a$:
    \[
    \rho_a = \frac{|POS_B(X)|}{|U|}
    \]
    \STATE Calculate importance $\gamma_a$:
    \[
    \gamma_a = \frac{|POS_B(X)| - |POS_{B-\{a\}}(X)|}{|U|}
    \]
    \STATE Compute overall weight $\omega_a$:
    \[
    \omega_a = \alpha \cdot \rho_a + \beta \cdot \gamma_a
    \]
\ENDFOR
\STATE Normalize overall weights $\omega_a'$:
\[
\omega_a' = \frac{\omega_a}{\sum_{a \in A} \omega_a}
\]
\STATE Screen samples based on the computed overall weights:
\[
X' = \{x \in U \mid \sum \omega_a' \cdot f(x, a) \geq \theta\}
\]
\STATE Output the screened sample set $X'$
\end{algorithmic}
\end{breakablealgorithm}

\section{A NOVEL CBT BASED CNSP TREATMENT SYSTEM}
We propose a novel interactive system for CNSP treatment. The first module demonstrates the detection of pain processes, including EMG and IMU, which jointly detect and record pain intervals. The second module shows the interactive treatment model of CBT embedded in the system and combines mobile devices with collaborative cognitive therapy to assist users in improving chronic pain. The following section examines these two main modules (pain detection and cognitive therapy) and the models designed through specific analyses. The system is shown below:

\begin{figure}[H]
    \centering
    \includegraphics[width=\columnwidth]{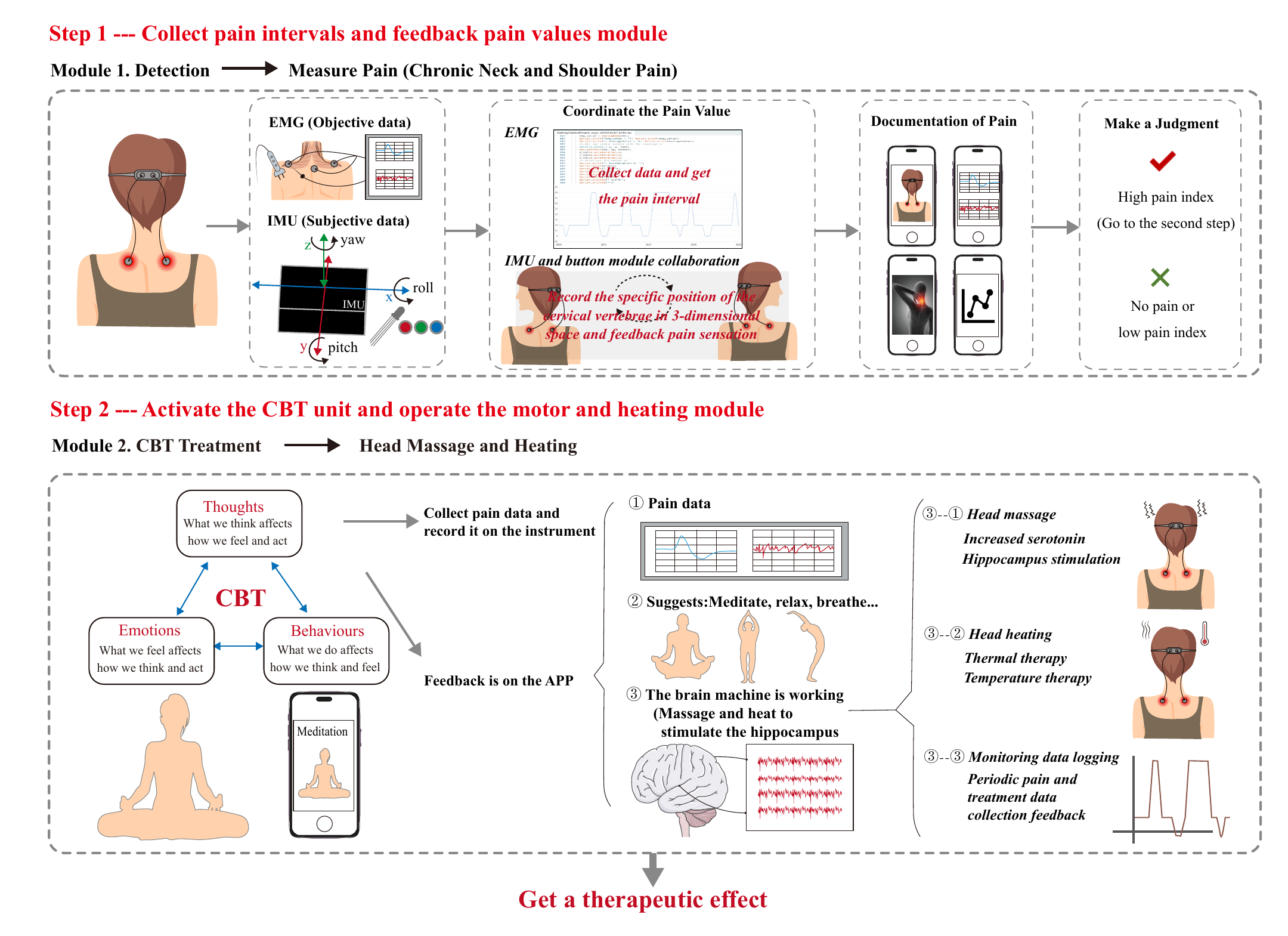}
    \caption{A novel interactive system for the treatment of CNSP. It is divided into two interactive steps. Step 1 is pain collection and recognition. Step 2 is CBT adjunctive treatment.}
    \label{fig2}
\end{figure}

\subsection{Pain Detection Modules}

The purpose of this integrative approach in our study is to improve the accuracy of pain management treatments by aligning physiological and UX-reported perspectives. The information architecture looks below:

\begin{figure}[H]
    \centering
    \includegraphics[width=\columnwidth]{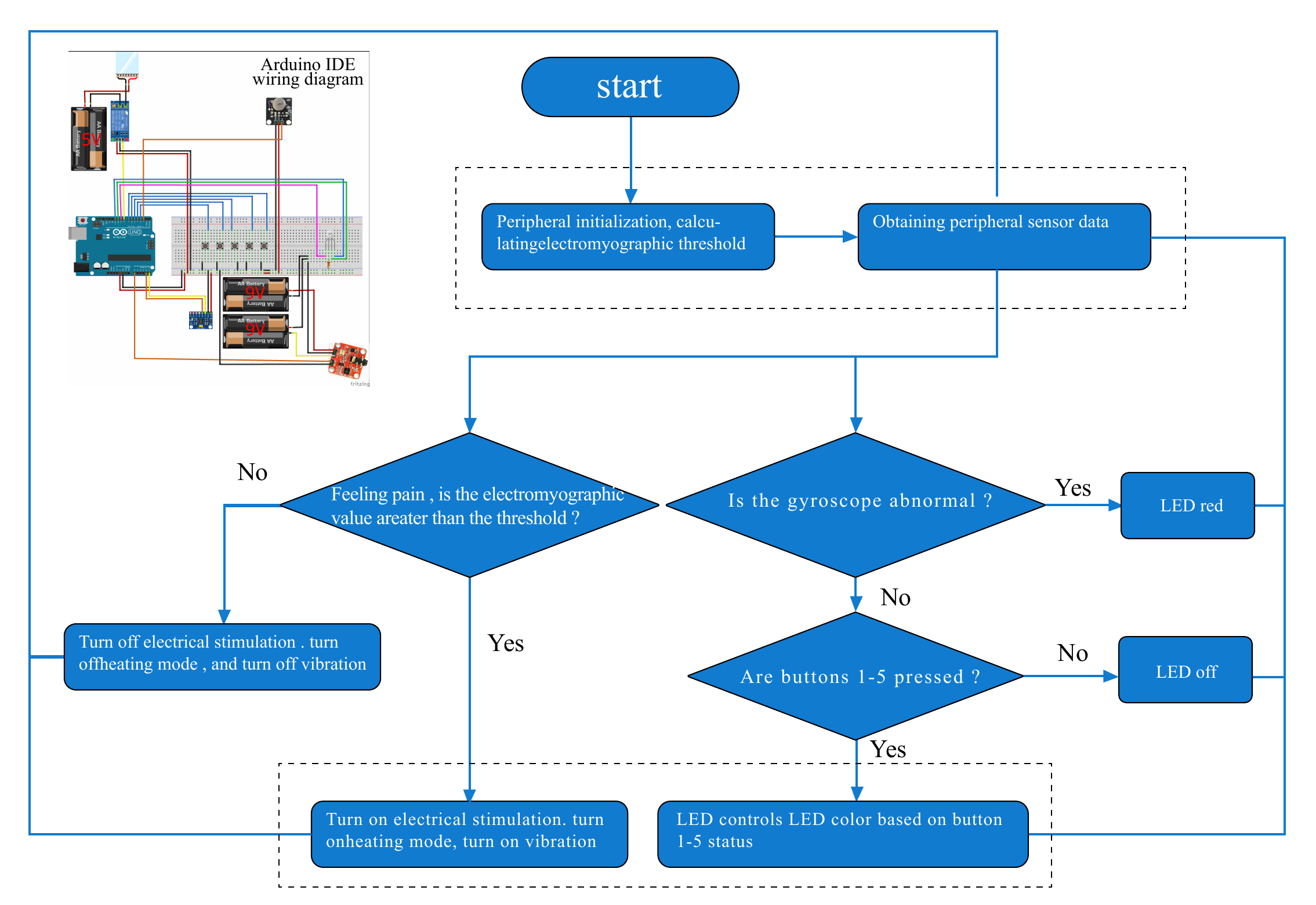}
    \caption{Pain detection Arduino sensor wiring logic diagram.}
    \label{fig3}
\end{figure}

Our system uses EMG pain testing to assess muscle and nervous system function objectively. EMG collects quantitative data to provide insight into the physiological aspects of pain. The EMG data collected is presented in terms of bit frequency, which is the regularity of the waveforms. Massive and tiny peaks are removed, and mean values are used to make the data specific to pain value intervals. Meanwhile, our system integrated an IMU and a button module to provide subjective pain feedback to patients. The aim is to facilitate real-time interactive communication, with the IMU capturing data on shoulder and neck trajectories and axial coordinates in 3D space. At the same time, the differently colored button modules allowed users to communicate and provide feedback on their pain experience with the mobile phone.

\subsection{Cognitive Therapy Modules and Hardware}
The hardware that enables CBT includes an Arduino dual-channel electrical stimulation board, medical-grade electrode pads, and a muscle stimulator. An auxiliary control electrode microcontroller ensures a precise combination of electrode pads and vibration frequencies, increasing user safety and therapeutic effectiveness \cite{b24}. 

CBT-assisted physiotherapy is a system that combines massage and heat. The choice of massage and heat comes from the patient's everyday use of pain-free massage and therapeutic strategies \cite{b14}. Massage promotes the unblocking of meridians, activates the hippocampus, and increases serotonin while heating to a specific temperature provides effective pain relief \cite{b19,b20}. The designed structure looks like the below prototypes:

\begin{figure}[H]
    \centering
    \includegraphics[width=\columnwidth]{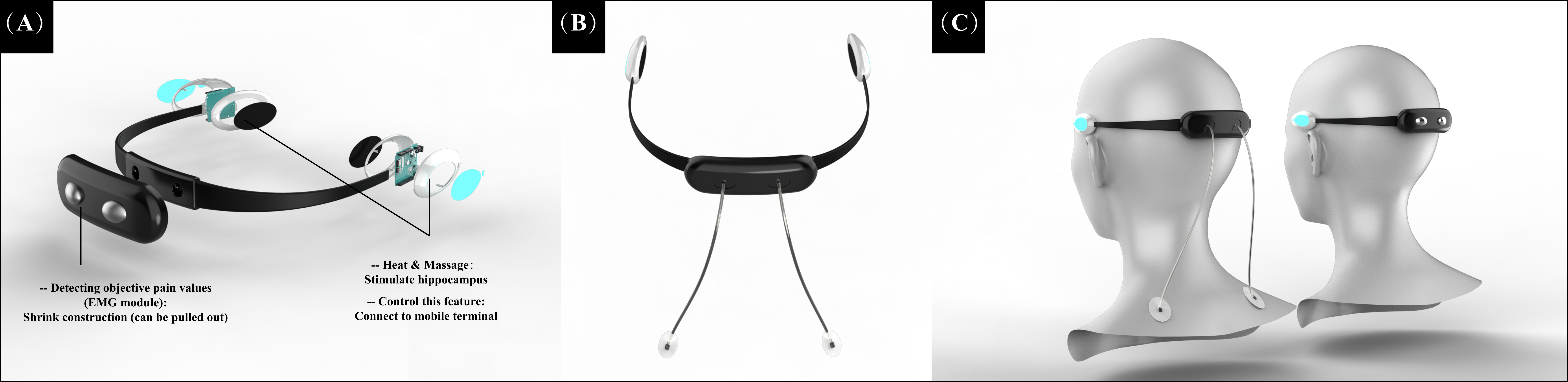}
    \caption{Equipment diagrams to assist in the treatment of CNSP. (A) Exploded view of equipment and location of sensors. (B) Panoramic view of head and neck apparatus. (C) Simulation of figure wearing.}
    \label{fig5}
\end{figure}

\section{EVALUATION PLAN}

\subsection{Experiment}
The exact use scenarios and interaction products are expected to be developed, and quantitative values of changes in nociception will be collected through a database of specified scales. Each group will consist of 15-20 participants to ensure equal numbers and the power of the data. This research approach will use qualitative and quantitative methods \cite{b6}. (1) Qualitative methods: We will use semi-structured face-to-face interviews. We aim to explore treatment outcomes from our study of patients with chronic pain \cite{b11}. (2) Quantitative methods: Questions to assess pain and psychological aspects were included in the questionnaire using a 5-point Likert scale. To ensure a comprehensive understanding of pain and psychological indicators \cite{b27}.

\subsubsection{Experimental Environment}
The equipment and location for this experiment will be in the same room, with the same setup, at the same time of day, with no other distractions \cite{b13,b37}. The experimental questionnaire will analyse the diversity of the research sample for valid sampling and data collection according to the topic's content.

\subsubsection{Participants}

To ensure diversity and comprehensiveness of data, participants in this study are people of any age, gender, or occupation with a history of chronic pain. This approach will capture different people's preferences and pain experiences when using the system. Participants could use the system anytime, anywhere, without needing a researcher or healthcare professional \cite{b3}. It also allowed participants to document and explore the functional design, for example, by creating schedules or generating tools as numerical comparisons of time intervals for pain physiotherapy.

\subsection{Evaluation Plan}
Qualitative research evaluations will subjectively analyze unstructured texts from interviews and observations. We will seek several experts to review and interpret the texts to ensure reliability independently \cite{b16}. We will also use thematic analysis using NVivo \cite{b41}.

Quantitative research evaluations will use quantifiable results to illuminate the objective reality of the CNSP treatment. We will use a 5-point Likert scale, one-way ANOVA, and a t-test for regular distribution to determine how significant the F and P values of the two groups are \cite{b38}. We will then compare each group's significance to ensure accurate data comparisons. The complete data will be statistically analyzed using SPSS.

\subsubsection{Quantitative Data Collection and Analysis - Schematic diagram of the IMU and button module synergizing to feedback the user's pain sensation}

\begin{figure}[H]
    \centering
    \includegraphics[width=\columnwidth]{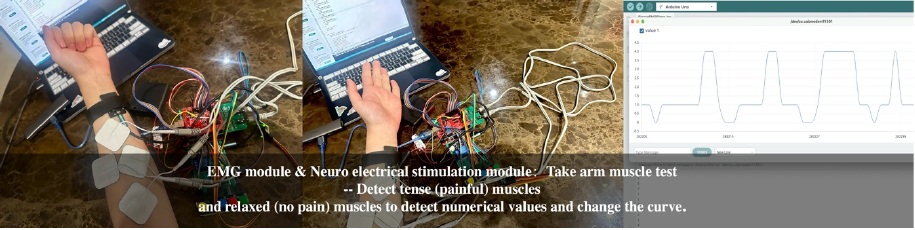}
    \caption{EMG Module and Electrical Nerve Stimulation Module: perform muscle testing to detect tense (painful) muscles and relaxed (pain-free) powers to detect values and capture pain intervals.}
    \label{fig4}
\end{figure}

The EMG detection module here is the one helping the user to proactively detect neck and shoulder pain, which we use the Arduino to directly detect to get the raw data, and later on use the rough set to do the data selection for the most important and relative data.

\begin{figure}[H]
    \centering
    \includegraphics[width=\columnwidth]{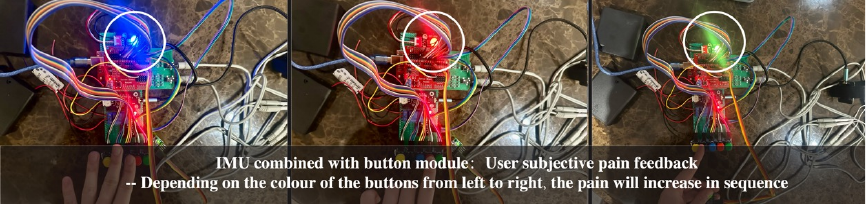}
    \caption{Add on quantitative data collection method, directly by user's input for pain sensation}
    \label{fig7}
\end{figure}

The IMU combined with the button's module is the one helping the user to directly input his/her feelings into the system by pressing the button, this method can be calculated as an add-on quantitative data collection method for reference.

\subsubsection{Multi sensor to improve effective of rough set}
In cases where there is only one attribute, the indiscernibility relation (\(IND(B)\)) becomes very intuitive. All pairs of data points that share the same value on this attribute are considered indistinguishable. However, all analyses will be limited to this single attribute. Nevertheless, using more sensors can enhance precision, and the indiscernibility relation will become more complex. In such scenarios, the use of rough set theory can be more effective.

\section{DISCUSSION}
\subsection{Analysis of Expected Quantitative and Qualitative CNSP Intervention Outcomes}

The system we studied is expected to be able to locate and relieve patients' pain in experiments. First, the pain detection module enables the system to monitor the user's neck and shoulder pain, providing a more personalized, real-time pain assessment. Second, adjunctive nociceptive therapy is based on CBT, which combines massage and heat functions to stimulate the hippocampus area of the user's brain and activate serotonin \cite{b4}. Taken together, this effect provides a more acceptable and practical treatment option for CNSP patients while alleviating the discomfort and anxiety associated with traditional treatments.

\subsection{Analysis of the Impact of Integrating Testing and Treatment within Metaverse}

The use of metaverse technology makes data collection and analysis during the treatment process more efficient and comprehensive. Various sensor technologies embedded in the virtual environment can seamlessly collect patient behavior data, physiological responses, and psychological feedback. This data is analyzed using big data and machine learning technologies, enabling real-time adjustments to the treatment plan to meet the patient's immediate needs and continuously optimize therapeutic outcomes over the long term. Through in-depth analysis of large datasets, researchers and doctors can gain a better understanding of the complexities of CNSP, advancing personalized medical services, improving treatment efficacy, and enhancing patient satisfaction.

\section{CONCLUSION}
This study developed a comprehensive system for EMG signal processing using Arduino-based hardware and rough set theory, divided into two main algorithms. Algorithm 1 involves initializing the Arduino system, collecting raw EMG samples, and processing them to calculate key features like average, minimum, and maximum values, with optional filtering and signal amplification. Algorithm 2 normalizes the processed EMG data and applies rough set theory to define an information system, calculate attribute importance, and screen samples to retain significant EMG features. This robust system enhances EMG signal analysis by addressing noise reduction, signal amplification, and feature selection, providing valuable insights into muscle activity in the evaluation plan. Future work aims to refine the algorithms and explore clinical applications, integrating advanced machine learning for real-time EMG analysis.

\end{document}